\documentclass{article}
\usepackage{graphicx} 

\usepackage[utf8]{inputenc}
\usepackage[T1]{fontenc}
\usepackage{lmodern}
\usepackage{amsmath,amssymb}
\usepackage{cite}
\usepackage{hyperref}
\usepackage{caption}
\usepackage{subcaption}
\usepackage{float}

\title{Detector-level assessment of alternative target nuclei for CEvNS experiments under realistic experimental conditions} 
\author{Yusuf Havvat\\[2mm]
\small Department of Physics, Cukurova University, 01330 Adana, Turkey\\
\small \texttt{havvatyusuf249@gmail.com}
}

\date{}

\begin{document}

\maketitle

\begin{abstract}
Coherent Elastic Neutrino--Nucleus Scattering (CEvNS) provides a sensitive probe of neutrino interactions at low momentum transfer, but its experimental observation is strongly constrained by detector-related effects such as energy threshold, resolution, noise, and event-selection criteria. In this work, we perform a detector-level assessment of CEvNS nuclear recoil observability under realistic experimental conditions, with particular emphasis on the role of detector response in shaping measurable recoil spectra. Using detailed \textsc{Geant4}-based simulations, CEvNS interactions are modeled for a set of alternative target nuclei spanning light to intermediate mass ranges. The true nuclear recoil energy distributions are propagated through a simplified yet realistic detector-response chain incorporating energy smearing, noise-induced fluctuations, threshold cuts, and veto-based event selection.

We present a systematic analysis of recoil energy spectra before and after detector effects, response matrices linking true and reconstructed energies, and energy-dependent selection efficiencies. The results demonstrate that detector response effects significantly modify the observable CEvNS signal, particularly in the near-threshold region where most recoil events are concentrated. Differences in efficiency turn-on behavior and reconstructed energy distributions highlight the target-nucleus dependence of CEvNS observability under identical detector conditions. Rather than focusing on absolute event-rate predictions, this study emphasizes the relative impact of detector effects on signal accessibility and target performance.

The presented framework provides a consistent methodology for evaluating and comparing prospective CEvNS target materials at the detector level, offering practical guidance for future low-threshold CEvNS experiments and detector design optimization.
\end{abstract}

\section{Introduction}

Coherent Elastic Neutrino--Nucleus Scattering (CEvNS) is a neutral-current weak interaction in which a low-energy neutrino scatters elastically from an entire nucleus, resulting in a coherently enhanced cross section that scales approximately with the square of the neutron number~\cite{Freedman1974CEvNS}. Owing to its coherent nature, CEvNS becomes dominant at low momentum transfer and produces nuclear recoils with kinetic energies typically ranging from a few tens of electronvolts to a few kiloelectronvolts, depending on the target nucleus and neutrino energy. Although the interaction probability is relatively large compared to other low-energy neutrino processes, the extremely small recoil energies make experimental detection highly challenging.

The experimental observation of CEvNS has only recently become feasible with the development of low-threshold detectors and advanced background-rejection techniques. The first unambiguous detection was achieved by the COHERENT collaboration using a CsI detector exposed to a stopped-pion neutrino source~\cite{COHERENT2017Science}. Subsequent measurements employing different target materials and detector technologies, including liquid argon and reactor-based germanium detectors, have further established CEvNS as a measurable process and a powerful probe of neutrino interactions and nuclear structure~\cite{COHERENT2021LArPRL,COHERENT2022CsICrossSection,CONUS2021PRLlimit,CONUS2024FinalPRL}. These experimental efforts have demonstrated that CEvNS sensitivity is strongly influenced by both target properties and detector performance.

From an experimental standpoint, the observability of CEvNS signals is determined not only by the underlying interaction cross section but also, critically, by detector-related effects. Finite energy resolution, electronic noise, energy thresholds, and event-selection criteria can significantly distort the measurable recoil-energy spectrum, particularly near threshold where the majority of CEvNS events are expected to occur~\cite{Scholberg2006Prospects}. As a result, theoretical recoil spectra or idealized rate estimates alone are insufficient to assess the practical detectability of CEvNS signals without incorporating a realistic detector-response model.

Monte Carlo simulations play a central role in bridging the gap between theoretical CEvNS predictions and experimental observables. Simulation frameworks based on \textsc{Geant4} provide a flexible and well-established platform for modeling particle interactions, detector geometry, and material properties at the detector level~\cite{Agostinelli2003Geant4,Allison2016Geant4}. Previous simulation studies have investigated CEvNS recoil spectra and expected event rates for a variety of target materials, often under simplified assumptions regarding detector response or selection efficiency~\cite{PapouliasKosmas2018PRD,BillardJohnstonKavanagh2018JCAP}. While these studies offer valuable insight into CEvNS phenomenology, they do not always capture the full impact of realistic detector effects on signal observability.

Motivated by these considerations, the present work focuses on a detector-level assessment of CEvNS observability under realistic experimental conditions. Rather than emphasizing absolute event-rate predictions, we investigate how detector response effects shape the measurable recoil-energy distributions and selection efficiencies for CEvNS signals. Particular attention is given to alternative target nuclei spanning light to intermediate mass ranges, which may offer complementary advantages for future CEvNS experiments when detector performance is taken into account.

Using detailed \textsc{Geant4}-based simulations, CEvNS-induced nuclear recoils are modeled and propagated through a simplified yet realistic detector-response chain that includes energy smearing, noise-induced fluctuations, threshold requirements, and veto-based event selection. By analyzing true and reconstructed recoil-energy spectra, response matrices, and energy-dependent efficiencies, we provide a consistent framework for comparing target-dependent CEvNS observability under identical detector assumptions. The methodology presented here aims to inform the design and optimization of future low-threshold CEvNS experiments by clarifying the interplay between target choice and detector response.

This work extends our previous detector-independent benchmark analysis of CEvNS recoil observables (arXiv:2602.04771), incorporating realistic neutrino spectra and detector-response modeling.

\section{Simulation Framework and Event Generation}

The numerical simulations presented in this work are performed using the \textsc{Geant4} Monte Carlo toolkit (version 11.2), which provides a well-established and extensively validated framework for particle transport and detector-level modeling in low-energy nuclear physics applications~\cite{Agostinelli2003Geant4,Allison2016Geant4}. The simulation is designed to model coherent elastic neutrino--nucleus scattering (CEvNS) interactions and to reconstruct experimentally relevant recoil observables for multiple target nuclei.

The primary objective of the simulation framework is to generate nuclear recoil energy distributions, detector response functions, and selection efficiencies in a unified and self-consistent manner, allowing for a direct comparison between different target materials under identical detector conditions.

\subsection{Physics Modeling}

CEvNS interactions are implemented by explicitly modeling the elastic scattering of neutrinos off target nuclei within the Standard Model framework. The interaction rate scales approximately with the square of the weak nuclear charge, resulting in an enhanced cross section for medium-to-heavy nuclei~\cite{Freedman1974CEvNS,DrukierStodolsky1984NCdetector}. Nuclear form factor effects are incorporated to account for coherence loss at higher momentum transfers, ensuring physical consistency across the full recoil energy range considered.

Electromagnetic and hadronic processes relevant to low-energy nuclear recoils are handled using the standard \textsc{Geant4} low-energy physics lists, which have been extensively benchmarked for sub-MeV nuclear interactions~\cite{Allison2016Geant4}.

\paragraph{Neutrino Energy Spectrum.}

The incident neutrino energy distribution is modeled according to the
Michel spectrum corresponding to electron neutrinos ($\nu_e$) produced
from muon decay at rest (DAR),
\[
\mu^{+} \rightarrow e^{+} + \nu_e + \bar{\nu}_\mu .
\]

The neutrino energy spectrum follows the functional form

\begin{equation}
f(E_\nu) \propto E_\nu^{2} \left(E_{\max} - E_\nu\right),
\end{equation}

where the kinematic endpoint is
$E_{\max} = 52.8~\mathrm{MeV}$.
Neutrino energies are sampled using an acceptance--rejection algorithm
over the interval $0 \leq E_\nu \leq 52.8~\mathrm{MeV}$.

This configuration approximates a stopped-pion neutrino source,
similar to that employed in the COHERENT experiment,
and provides a realistic low-energy neutrino spectrum for CE$\nu$NS
studies.

\subsection{Target Materials and Event Statistics}

Four target nuclei---boron (B), magnesium (Mg), titanium (Ti), and zirconium (Zr)---are investigated in this study. \textbf{To isolate genuine mass-dependent effects, the simulations assume a single dominant isotope for each element (specifically $^{11}$B, $^{24}$Mg, $^{48}$Ti, and $^{90}$Zr) rather than natural isotopic mixtures.} Each target material is simulated independently using identical detector geometry, physics configurations, and reconstruction settings.

For each material, a sufficiently large ensemble of primary neutrino interactions is generated to ensure stable recoil-energy distributions and statistically meaningful detector-response observables. 
Due to intrinsic differences in nuclear mass, recoil kinematics, and reconstruction behavior, the number of reconstructed recoil events naturally varies among the targets. 
Heavier nuclei, benefiting from coherent enhancement, tend to yield larger reconstructed samples under identical simulation conditions, whereas lighter nuclei produce comparatively smaller but kinematically extended recoil populations.

All statistical variations are explicitly propagated into the response matrices and efficiency curves presented in subsequent sections. 
The relative comparisons between targets therefore reflect genuine detector-level and nuclear-physics differences rather than artifacts of finite Monte Carlo statistics.

\subsection{Detector Response Model}

A simplified yet realistic detector-response model is implemented
to translate true nuclear recoil energies into experimentally
reconstructed observables.

\paragraph{Energy Resolution.}

Energy smearing is modeled using a Gaussian response with
energy-dependent resolution,

\begin{equation}
\sigma(E) = \sqrt{a^{2}E + b^{2}},
\end{equation}

where $E$ is expressed in keV.
The resolution parameters are chosen as

\[
a = 0.20~\sqrt{\mathrm{keV}}, 
\qquad
b = 0.10~\mathrm{keV}.
\]

This functional form incorporates both a stochastic term
($\propto \sqrt{E}$) and a constant baseline contribution,
providing a realistic representation of low-threshold detector
performance.

\paragraph{Electronics Noise.}

An independent Gaussian electronics noise term with width

\[
\sigma_{\mathrm{noise}} = 0.2~\mathrm{keV}
\]

is added to the smeared recoil energy.

\paragraph{Energy Threshold and Veto Condition.}

A hard detection threshold of

\[
E_{\mathrm{thr}} = 1.0~\mathrm{keV}
\]

is applied to the reconstructed recoil energy.
In addition, events depositing more than

\[
E_{\mathrm{veto}} = 0.2~\mathrm{keV}
\]

in the veto detector volume are rejected.
The final event-selection variable is defined by requiring
both the threshold and veto conditions to be satisfied.

\subsection{Recoil Energy Definition}

Two distinct recoil energy observables are defined:

\begin{itemize}
  \item \textbf{True recoil energy ($E_{\mathrm{true}}$):}  
  The physical nuclear recoil energy directly obtained from the CEvNS interaction vertex prior to any detector effects.

  \item \textbf{Measured recoil energy ($E_{\mathrm{meas}}$):}  
  The reconstructed recoil energy after applying detector response effects, including energy threshold, resolution smearing, and veto conditions.
\end{itemize}

This separation allows for a transparent investigation of detector-induced distortions and their impact on the observable recoil spectrum.

\subsection{Output Observables}

For each target nucleus, the simulation produces the following observables:
\begin{enumerate}
  \item True nuclear recoil energy spectrum.
  \item Measured recoil energy spectrum after detector effects.
  \item Detector response matrix correlating $E_{\mathrm{true}}$ and $E_{\mathrm{meas}}$.
  \item Event selection efficiency as a function of recoil energy.
  
\end{enumerate}

These observables form the basis of the comparative analysis presented in Section 3, where the detector-level implications of different CEvNS target materials are discussed in detail.

\section{Results}
\label{sec:results}

In this section, the detector-level observability of CEvNS nuclear recoils is investigated through a comparative analysis of multiple target nuclei under identical simulation and detector-response conditions. The results are organized according to recoil observables rather than individual target materials in order to facilitate a direct and unbiased comparison of target-dependent effects. For each observable, the corresponding distributions are presented simultaneously for all targets, ensuring that differences arise solely from nuclear properties and detector response rather than from variations in simulation settings.

The analysis focuses on five key observables that characterize the experimental accessibility of CEvNS signals: the true nuclear recoil energy spectrum, the reconstructed recoil energy spectrum after detector effects, the detector response matrix linking true and measured energies, the impact of event-selection criteria, and the resulting detection efficiency as a function of recoil energy. Together, these observables provide a comprehensive detector-level description of CEvNS signal formation and enable a systematic evaluation of target performance under realistic experimental conditions~\cite{Scholberg2006Prospects,PapouliasKosmas2018PRD,BillardJohnstonKavanagh2018JCAP}.

Throughout this section, emphasis is placed on relative trends and target-dependent behavior rather than absolute event-rate predictions. This approach isolates the influence of detector response and nuclear recoil kinematics on signal observability and avoids reliance on experiment-specific neutrino fluxes or exposure assumptions. The results presented here therefore offer a robust and general framework for comparing prospective CEvNS target materials from an experimental standpoint.

\begin{figure}[H]
\centering

{\Large \textbf{True Nuclear Recoil Energy Spectra}}\\[0.5cm]

\begin{subfigure}{0.48\textwidth}
    \centering
    \includegraphics[width=\linewidth]{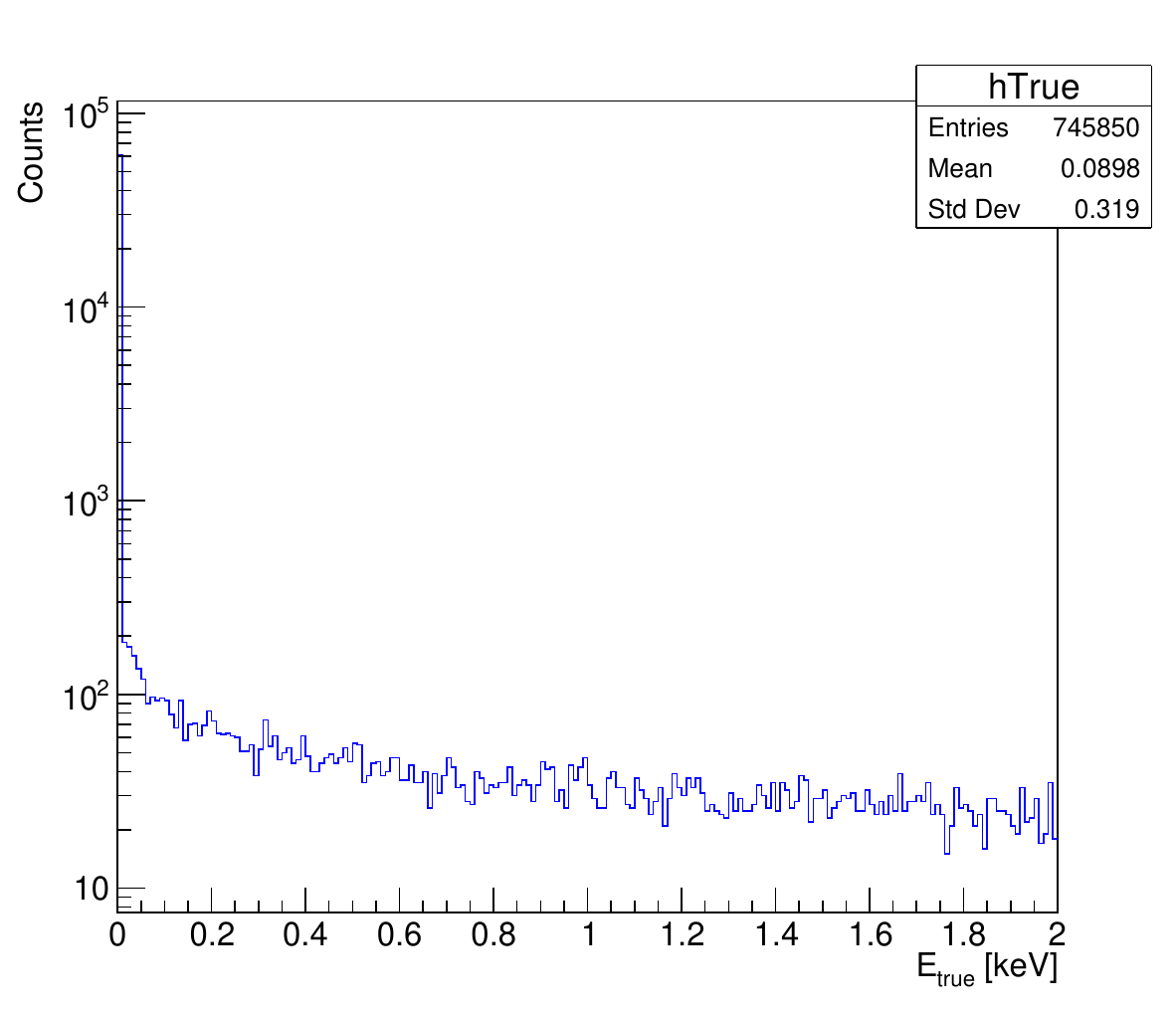}
    \caption{Boron (B) true recoil energy spectrum.}
\end{subfigure}
\hfill
\begin{subfigure}{0.48\textwidth}
    \centering
    \includegraphics[width=\linewidth]{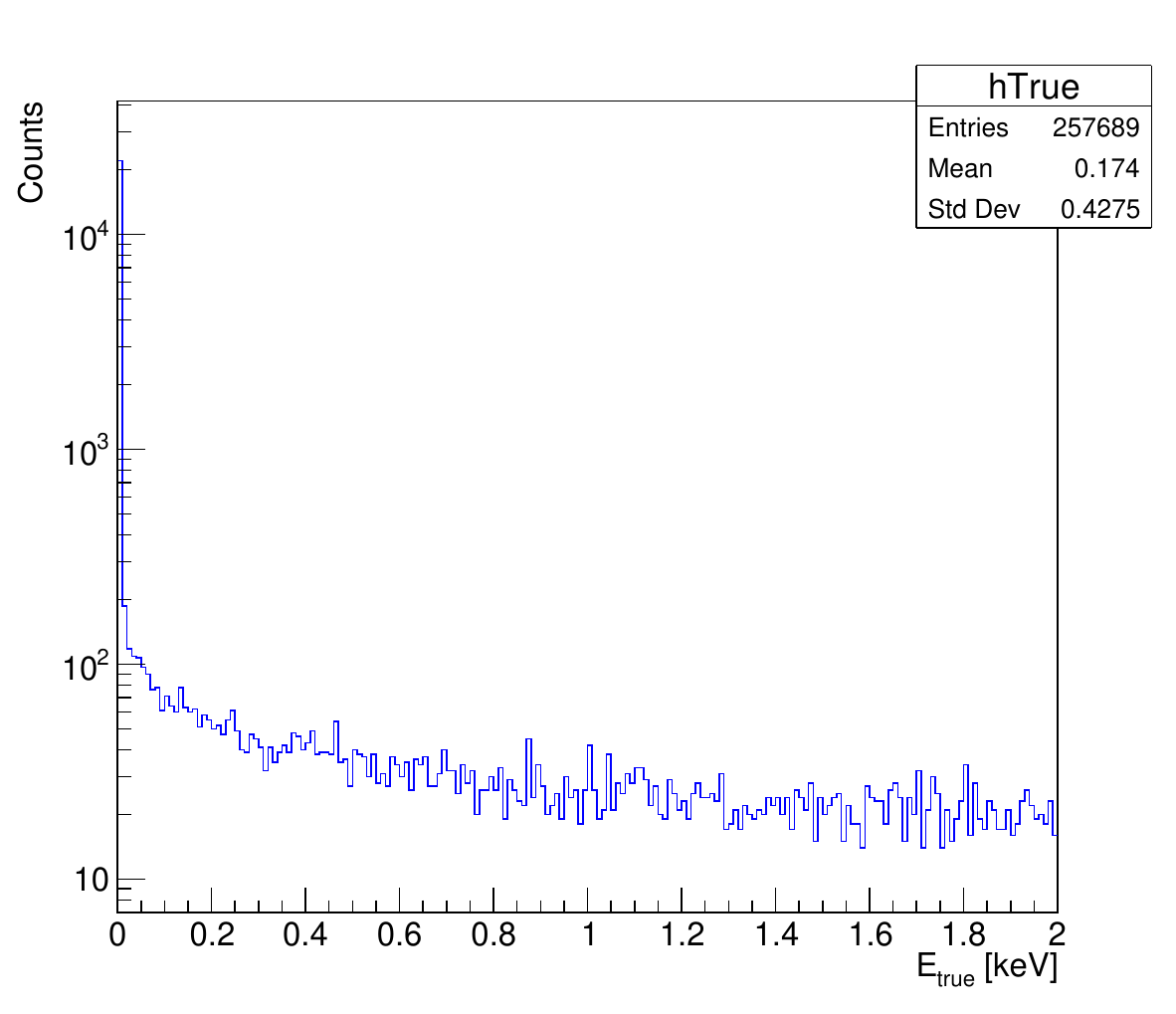}
    \caption{Magnesium (Mg) true recoil energy spectrum.}
\end{subfigure}

\vspace{0.5cm}

\begin{subfigure}{0.48\textwidth}
    \centering
    \includegraphics[width=\linewidth]{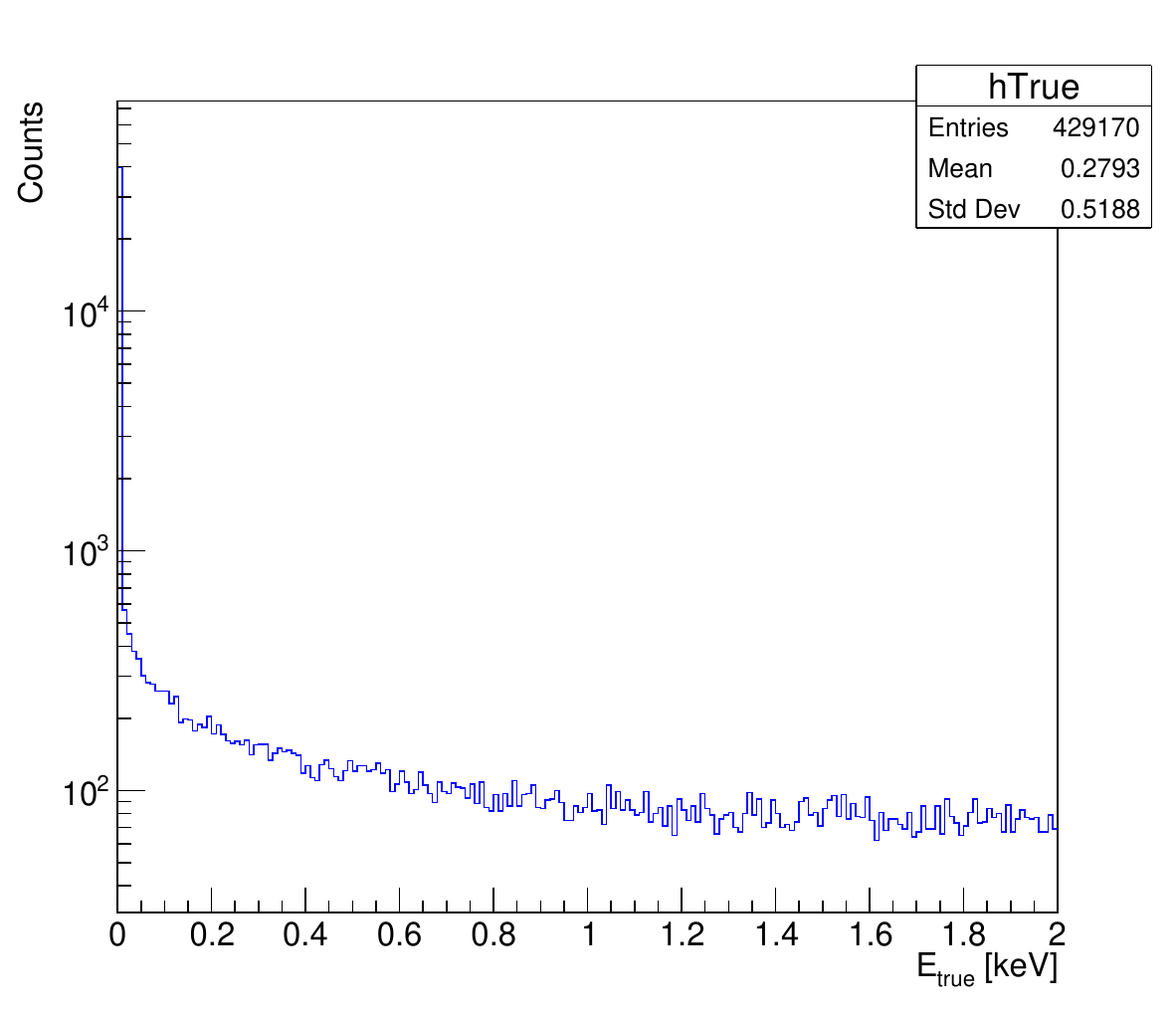}
    \caption{Titanium (Ti) true recoil energy spectrum.}
\end{subfigure}
\hfill
\begin{subfigure}{0.48\textwidth}
    \centering
    \includegraphics[width=\linewidth]{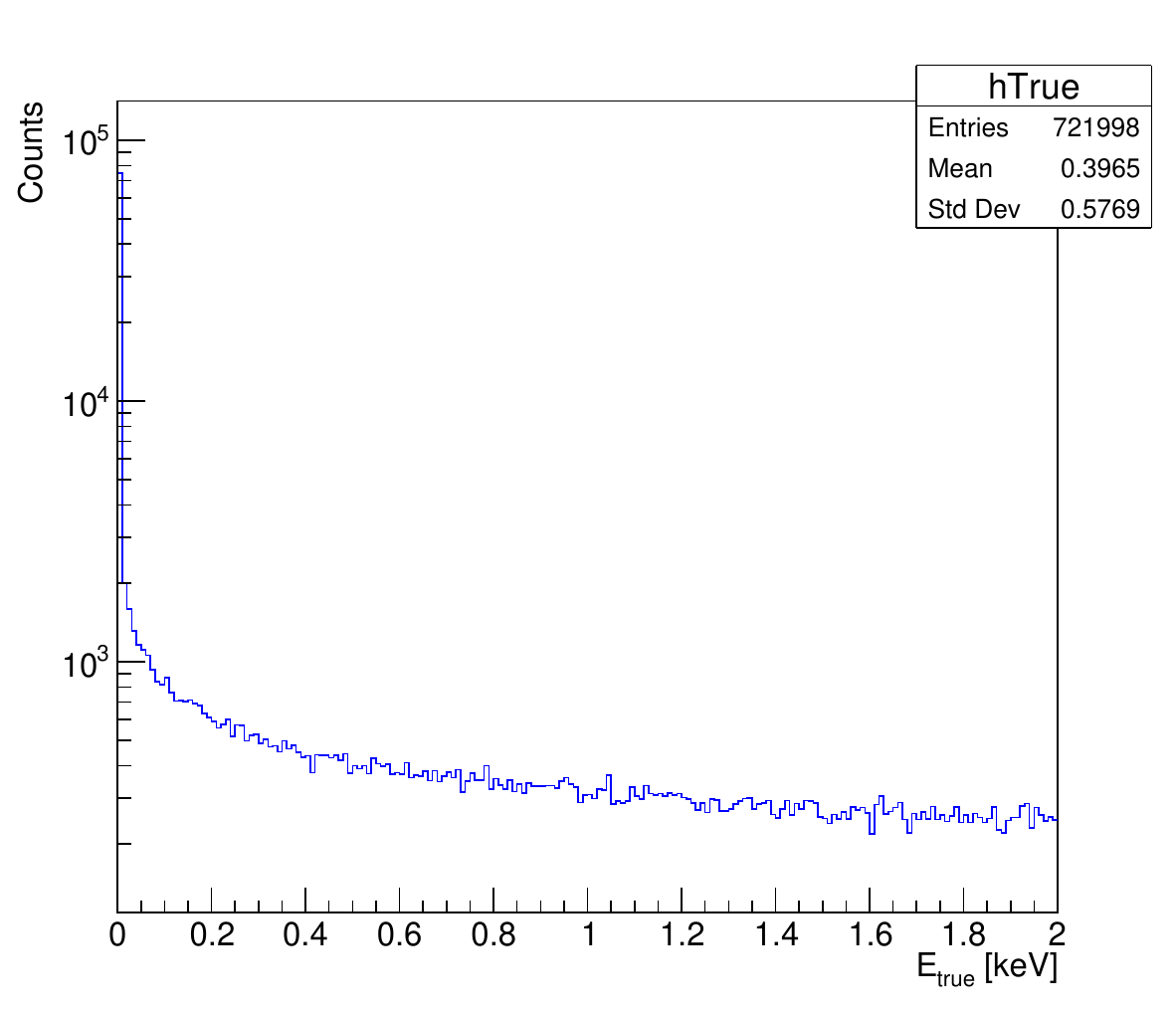}
    \caption{Zirconium (Zr) true recoil energy spectrum.}
\end{subfigure}

\caption{True nuclear recoil energy spectra ($E_{\mathrm{true}}$) for the four target nuclei, produced under identical simulation settings. No detector-response model (smearing, threshold, veto, or additional analysis cuts) is applied in this figure; the distributions therefore represent the underlying kinematic recoil-energy spectra prior to detector effects.}
\label{fig:etrue}
\end{figure}

\subsection{True nuclear recoil energy spectra}
\label{sec:true_energy}

Figure~\ref{fig:etrue} presents the true nuclear recoil energy spectra ($E_{\mathrm{true}}$) obtained from high-statistics CE$\nu$NS simulations for boron (B), magnesium (Mg), titanium (Ti), and zirconium (Zr) targets. Since no detector effects are included at this stage, these distributions represent the purely kinematic response governed by neutrino–nucleus elastic scattering.

For boron (Z=5), the recoil-energy distribution is strongly concentrated in the sub-keV region, with a mean recoil energy of $\langle E_{\mathrm{true}}\rangle \approx 0.0898$~keV and a broad relative spread. The spectrum exhibits a steeply falling shape characteristic of CE$\nu$NS scattering at low momentum transfer, where the differential cross section scales approximately as $d\sigma/dT \propto N^2$ with kinematic suppression at higher recoil energies. The dominance of events below $\sim 0.5$~keV indicates that light targets primarily probe the lowest recoil regime, making them intrinsically sensitive to detector threshold performance.

Magnesium (Z=12) displays a noticeable upward shift in recoil-energy scale, with $\langle E_{\mathrm{true}}\rangle \approx 0.174$~keV and a broader high-energy tail. The increased nuclear mass modifies the recoil kinematics such that the spectrum extends to higher recoil energies while still maintaining significant population in the sub-keV region. This intermediate-mass behavior reflects a balance between recoil reach and total cross section, consistent with the expected CE$\nu$NS scaling relations.

For titanium (Z=22) and zirconium (Z=40), the recoil spectra shift further toward higher energies, with visibly extended high-energy tails. Although heavier nuclei produce lower maximum recoil energies for a fixed neutrino energy in the extreme kinematic limit, the interplay between nuclear mass, neutrino spectrum, and form-factor suppression redistributes the recoil population toward a region that is less concentrated near zero. Consequently, the event density at very low energies is relatively reduced compared to lighter targets, while the distribution becomes more uniform across the accessible recoil range.

Importantly, these true spectra highlight the fundamental mass-dependent trade-off in CE$\nu$NS target selection. Light nuclei maximize recoil energy reach but concentrate events near threshold, increasing sensitivity to detector limitations. Heavier nuclei, by contrast, provide enhanced coherent cross section ($\propto N^2$) and a more experimentally favorable recoil-energy distribution once realistic detector effects are folded in. This mass-dependent spectral behavior underlies current experimental strategies employing CsI, liquid argon, and germanium detectors in CE$\nu$NS measurements.

Therefore, the true recoil spectra already encode the essential physics drivers of detector-level performance: nuclear mass, coherent enhancement, recoil kinematics, and spectral concentration in the threshold-dominated regime. Subsequent sections will demonstrate how these intrinsic differences propagate once realistic detector response and selection efficiencies are applied.

\begin{figure}[H]
\centering

{\Large \textbf{CEvNS Measured Recoil Energy Spectra ($E_{\mathrm{meas}}$)}}\\[0.5cm]

\begin{subfigure}[t]{0.48\textwidth}
    \centering
    \includegraphics[width=\linewidth]{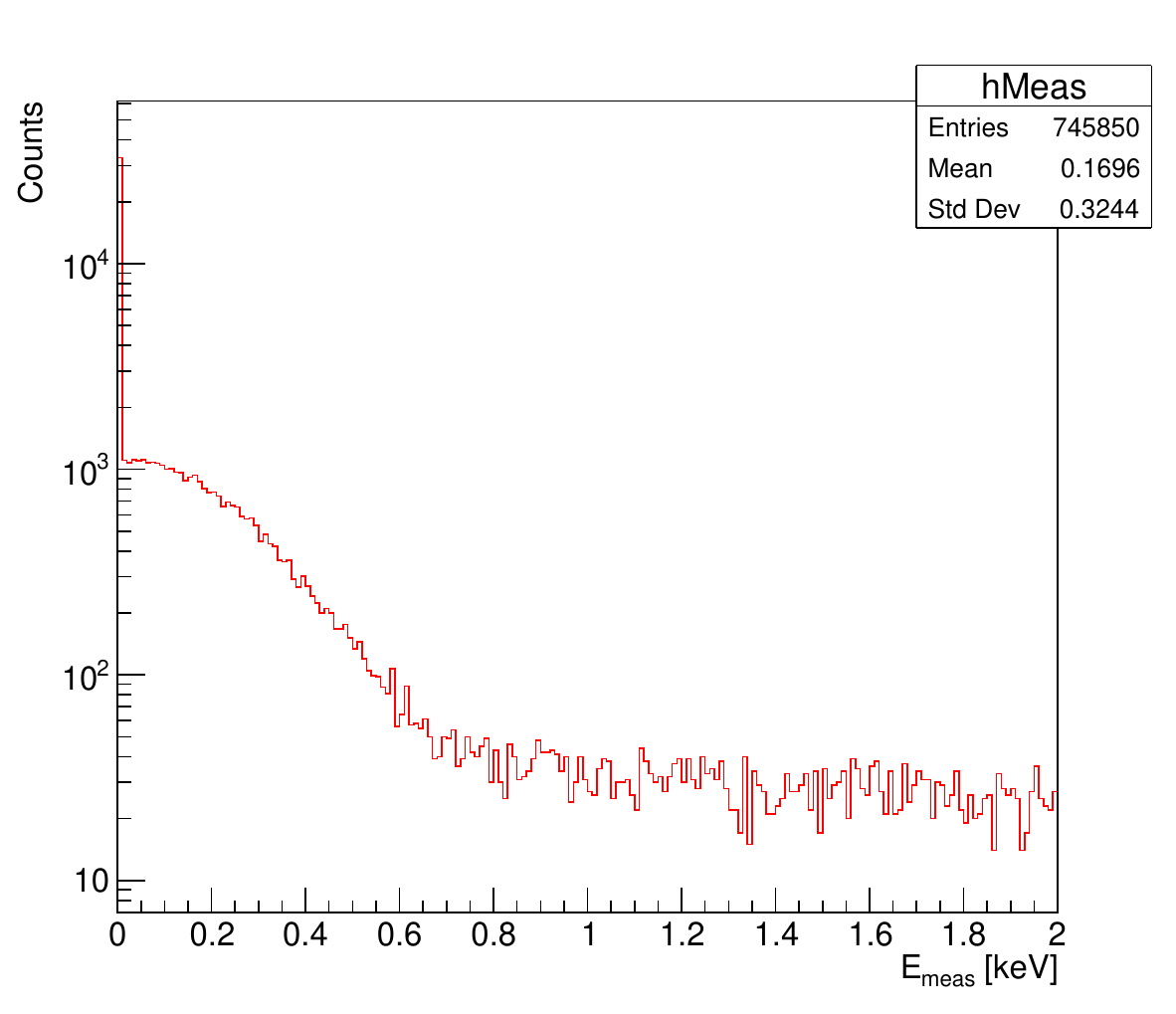}
    \caption{Boron (B) measured recoil energy spectrum.}
    \label{fig:emeas_B}
\end{subfigure}
\hfill
\begin{subfigure}[t]{0.48\textwidth}
    \centering
    \includegraphics[width=\linewidth]{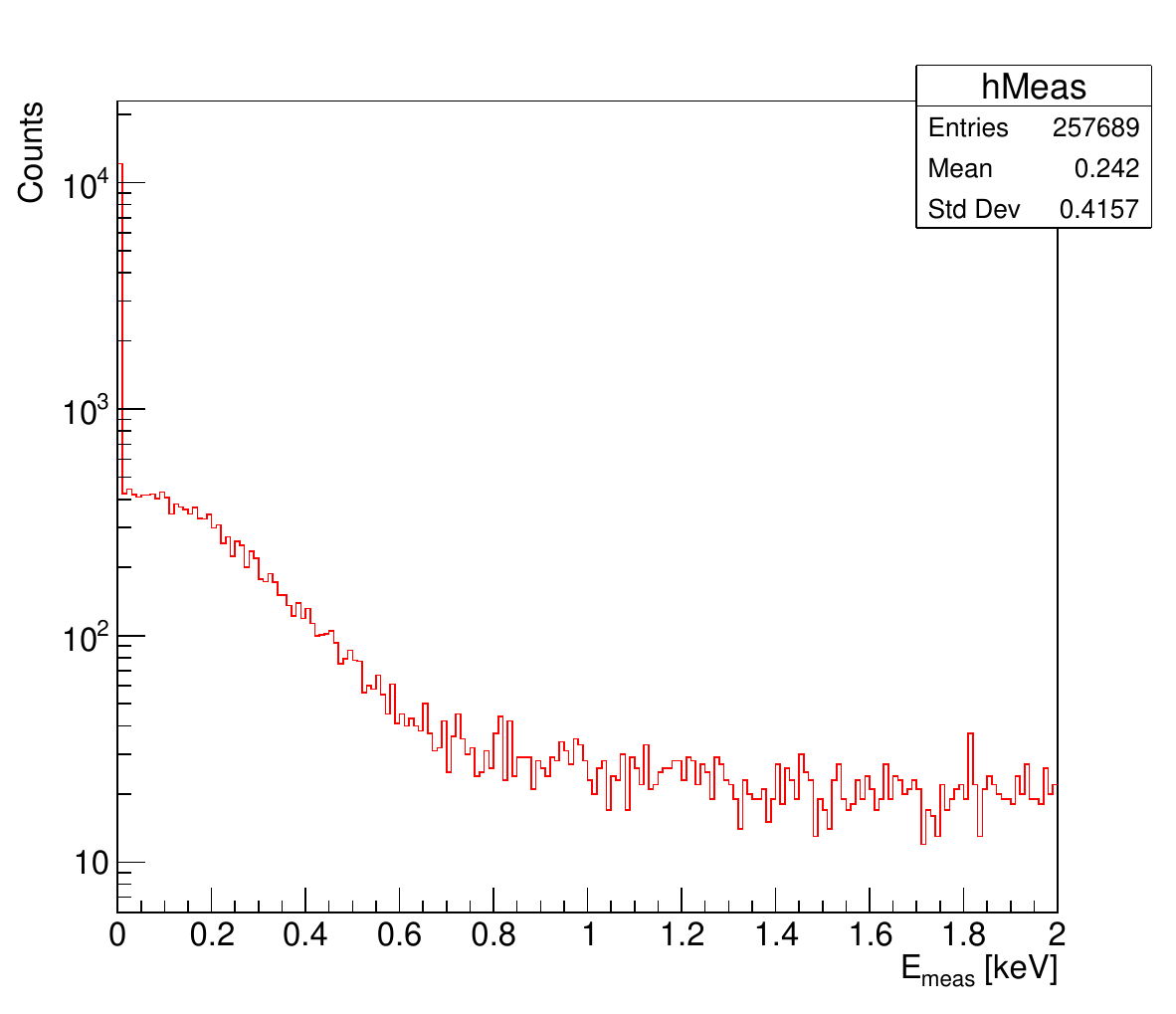}
    \caption{Magnesium (Mg) measured recoil energy spectrum.}
    \label{fig:emeas_Mg}
\end{subfigure}

\vspace{0.5cm}

\begin{subfigure}[t]{0.48\textwidth}
    \centering
    \includegraphics[width=\linewidth]{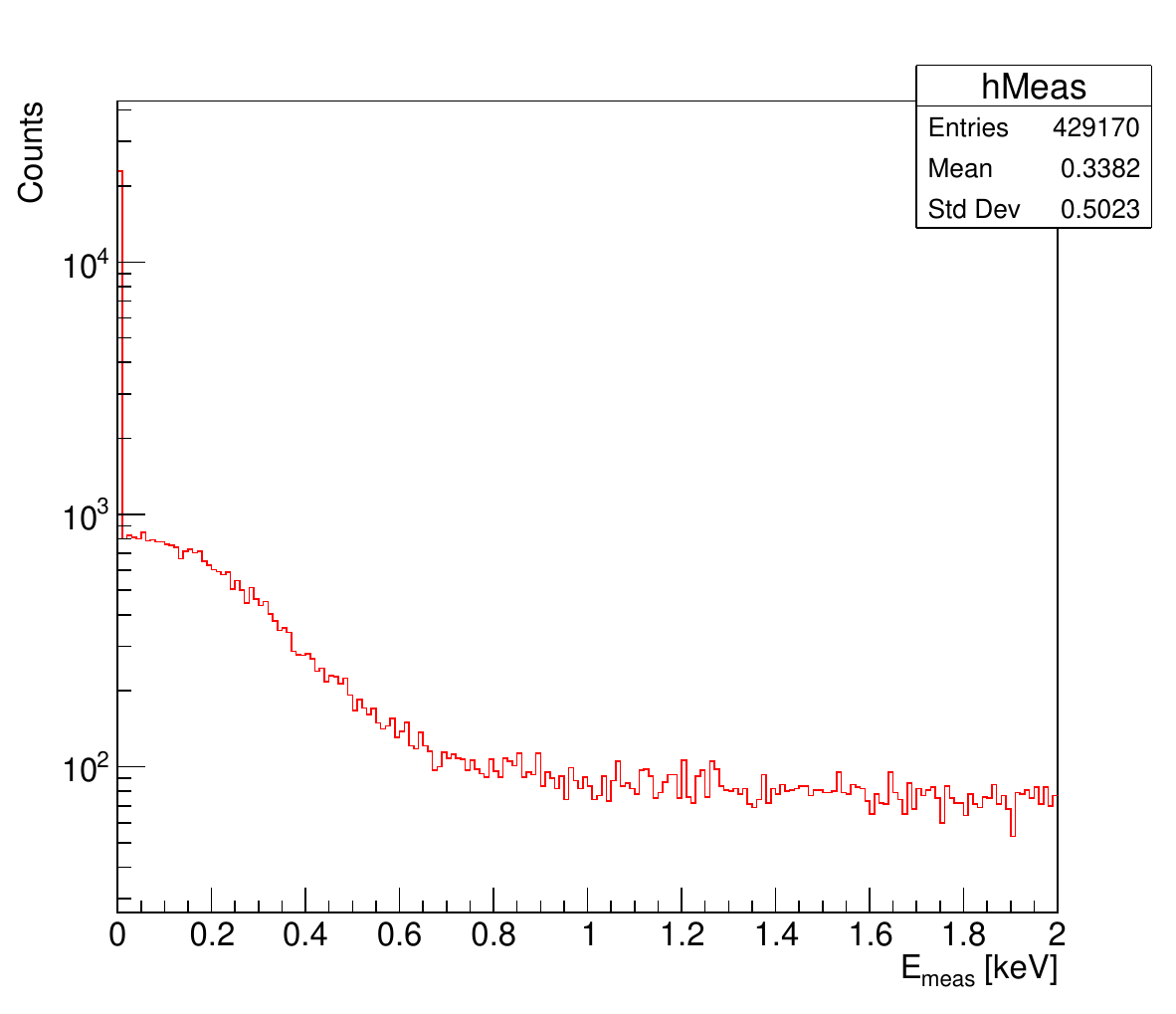}
    \caption{Titanium (Ti) measured recoil energy spectrum.}
    \label{fig:emeas_Ti}
\end{subfigure}
\hfill
\begin{subfigure}[t]{0.48\textwidth}
    \centering
    \includegraphics[width=\linewidth]{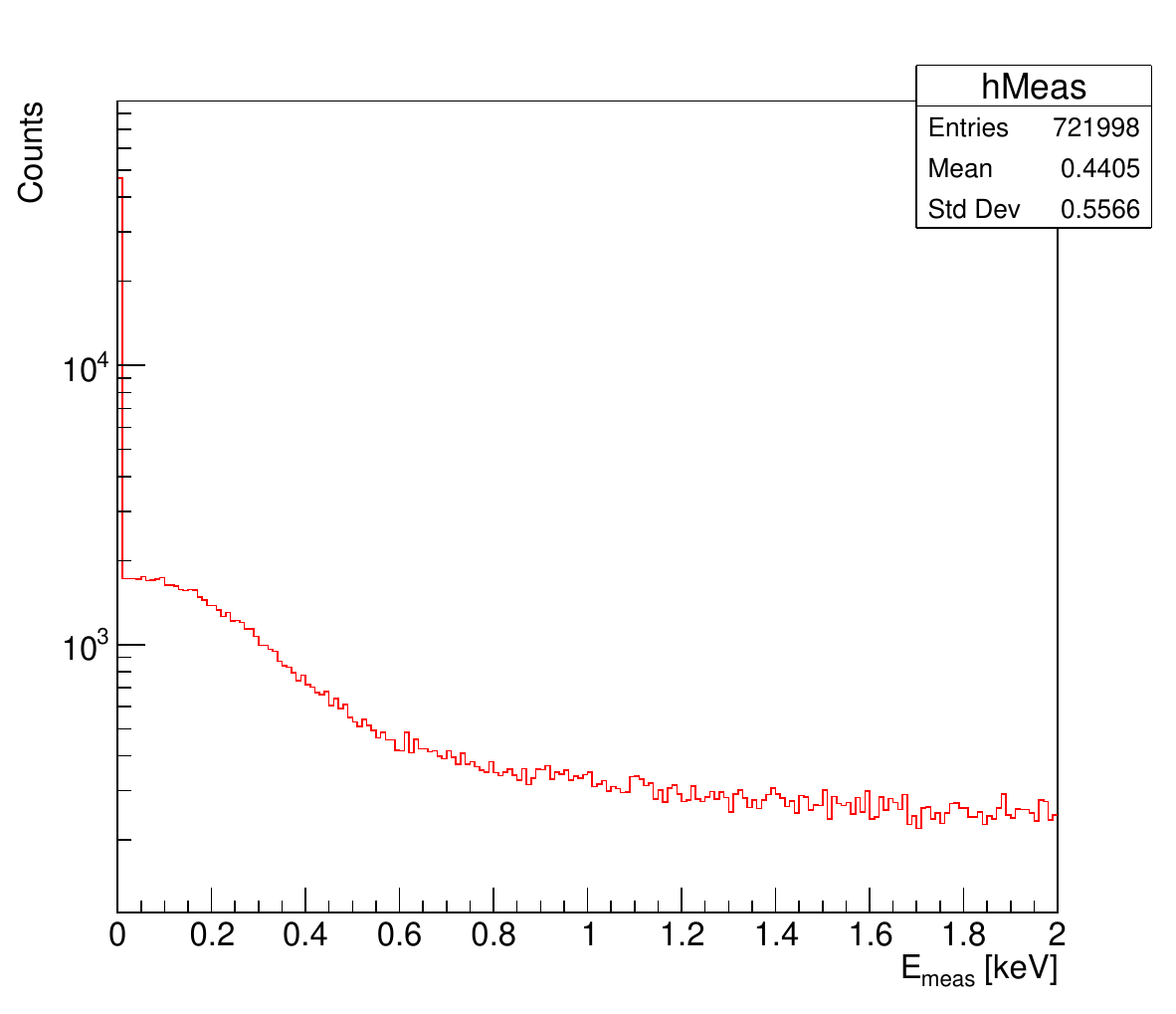}
    \caption{Zirconium (Zr) measured recoil energy spectrum.}
    \label{fig:emeas_Zr}
\end{subfigure}

\caption{Measured nuclear recoil energy spectra ($E_{\mathrm{meas}}$) obtained after applying detector response effects, including energy resolution, threshold, and event selection cuts, for the four investigated CEvNS target nuclei: (a) boron (B), (b) magnesium (Mg), (c) titanium (Ti), and (d) zirconium (Zr). All spectra are shown under identical simulation and analysis conditions, illustrating how detector effects reshape the experimentally accessible recoil energy distributions depending on target mass.}
\label{fig:emeas}
\end{figure}

Figure~\ref{fig:emeas} presents the detector-level reconstructed recoil energy spectra ($E_{\text{meas}}$) for B, Mg, Ti, and Zr targets under realistic experimental conditions including energy resolution and electronic noise effects. Unlike the true recoil distributions, these spectra incorporate detector response smearing, which is critical for assessing experimental observability.

For boron, the measured spectrum exhibits a strong accumulation below 0.5~keV, with a mean reconstructed energy of approximately 0.1696~keV. The sharp peak near threshold indicates that although boron benefits from kinematic enhancement at low recoil energies, a substantial fraction of events lies close to the detection limit. This behavior implies that boron-based CEvNS detection would require sub-keV threshold capability similar to silicon CCD-based technologies as demonstrated in the CONNIE experiment~\cite{CONNIE2019DataSearch}.

Magnesium shows a broader distribution with a higher mean reconstructed energy (~0.242~keV), indicating improved energy separation from threshold effects. The increased recoil energy relative to boron reduces threshold-induced efficiency losses while maintaining relatively high event statistics. This intermediate behavior suggests magnesium as a potential compromise between light and medium-mass nuclei in CEvNS detector optimization.

Titanium and zirconium demonstrate further shifts of the reconstructed spectrum toward higher recoil energies. The broader distribution and reduced threshold compression reflect the increasing nuclear mass, which enhances coherent scattering cross section proportional to $N^2$ while moderately reducing recoil amplitude. In particular, zirconium provides a balance between event yield and detector-level visibility, producing a spectrum less dominated by threshold effects compared to lighter nuclei.

When compared with current experimental efforts such as COHERENT’s CsI and LAr detectors~\cite{COHERENT2017Science,COHERENT2021LArPRL}, the reconstructed spectra of Ti and Zr fall within an experimentally accessible recoil window, whereas B and Mg emphasize the importance of ultra-low threshold detection technologies. 

The measured spectra clearly demonstrate that detector response effects significantly reshape the true recoil distributions. Therefore, target selection for future CEvNS experiments cannot rely solely on theoretical cross-section scaling but must incorporate realistic detector-level reconstruction. The observed differences among the four candidate nuclei highlight how nuclear mass and detector response interplay to determine experimental sensitivity.

\begin{figure}[H]
\centering
{\Large \textbf{Detector response matrix ($E_{\mathrm{meas}}$ vs $E_{\mathrm{true}}$)}}\\[0.5cm]

\begin{subfigure}{0.48\textwidth}
    \centering
    \includegraphics[width=\linewidth]{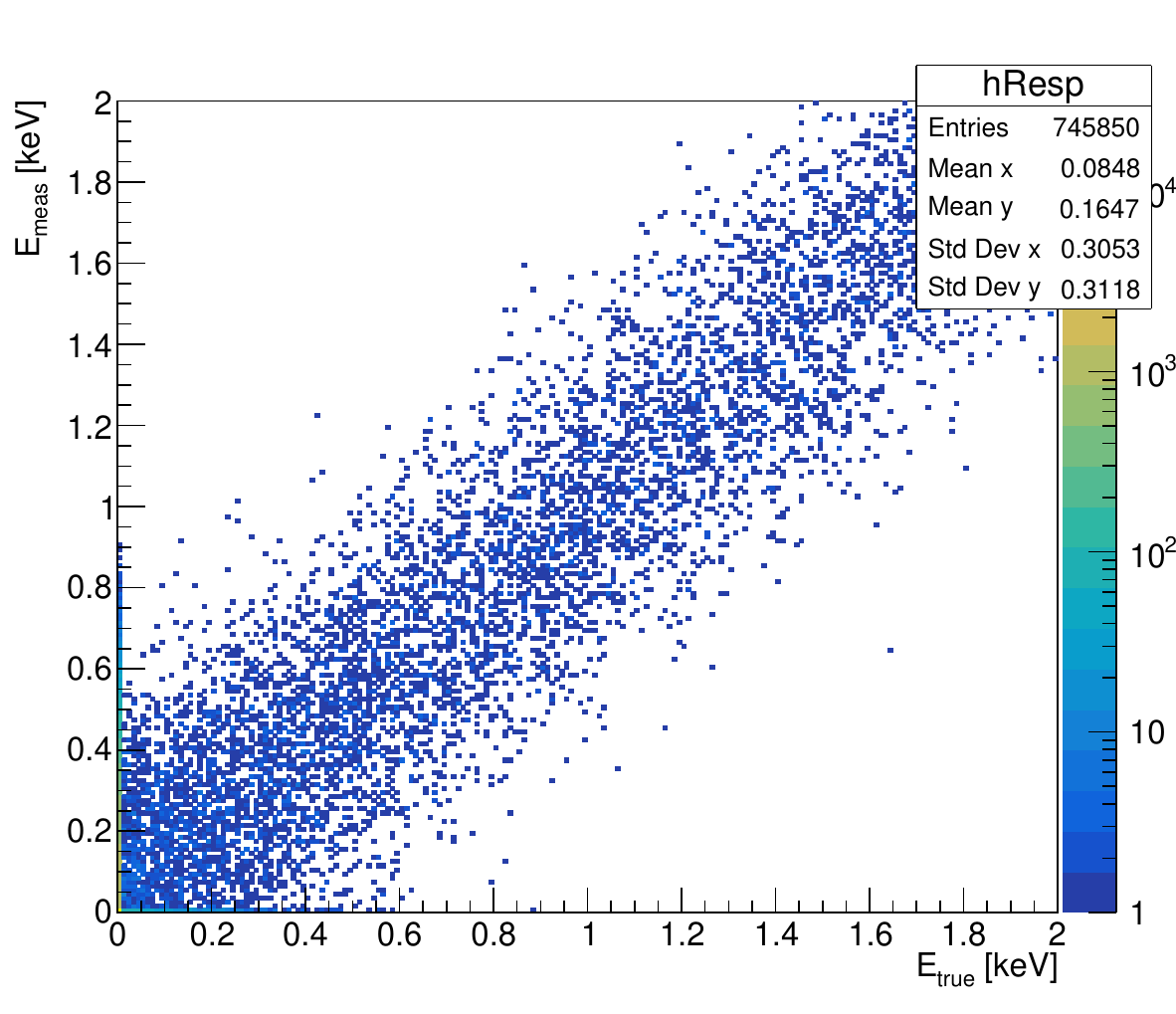}
    \caption{Boron (B): detector response matrix.}
    \label{fig:matrix_B}
\end{subfigure}
\hfill
\begin{subfigure}{0.48\textwidth}
    \centering
    \includegraphics[width=\linewidth]{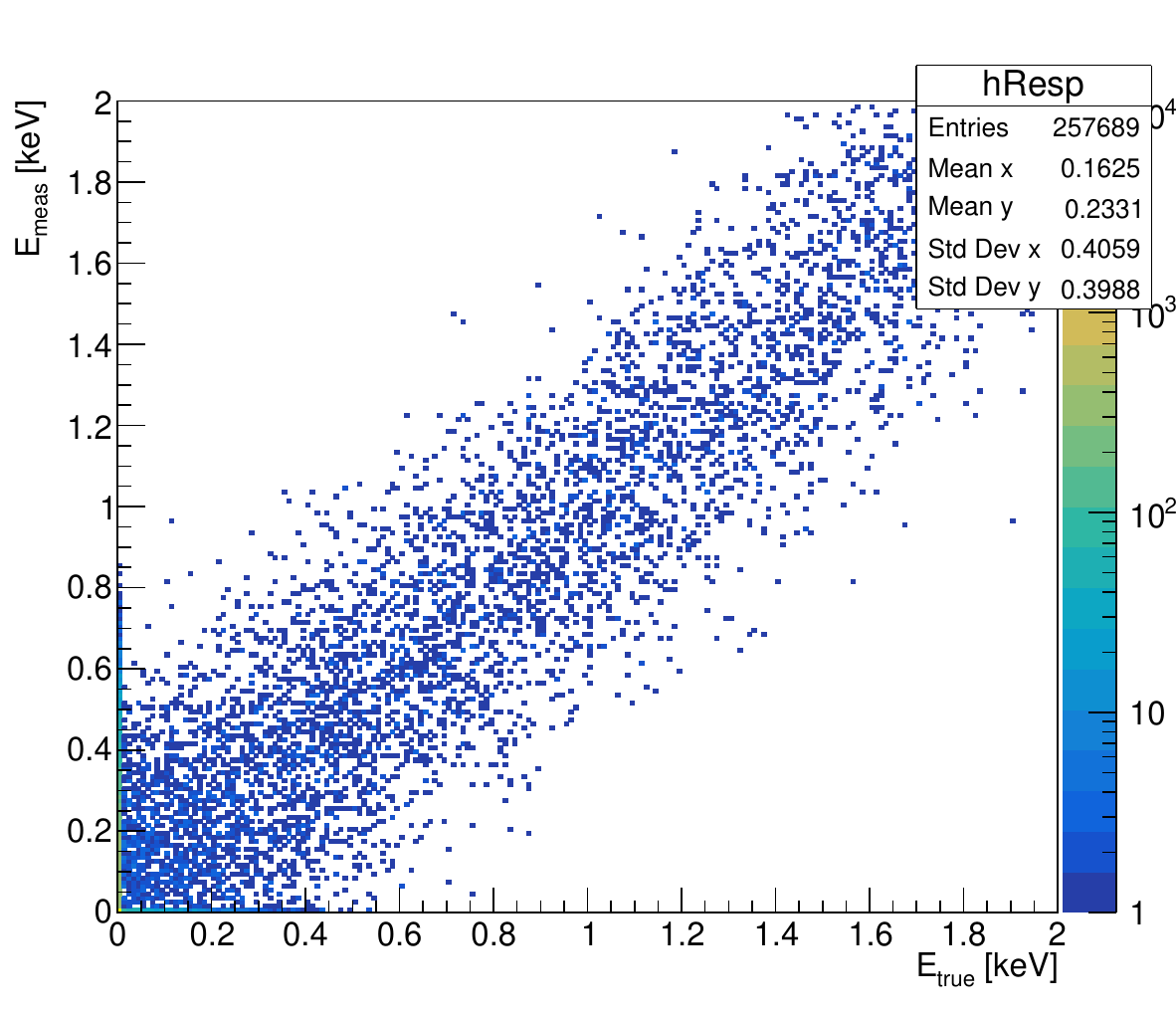}
    \caption{Magnesium (Mg): detector response matrix.}
    \label{fig:matrix_Mg}
\end{subfigure}

\vspace{0.5cm}

\begin{subfigure}{0.48\textwidth}
    \centering
    \includegraphics[width=\linewidth]{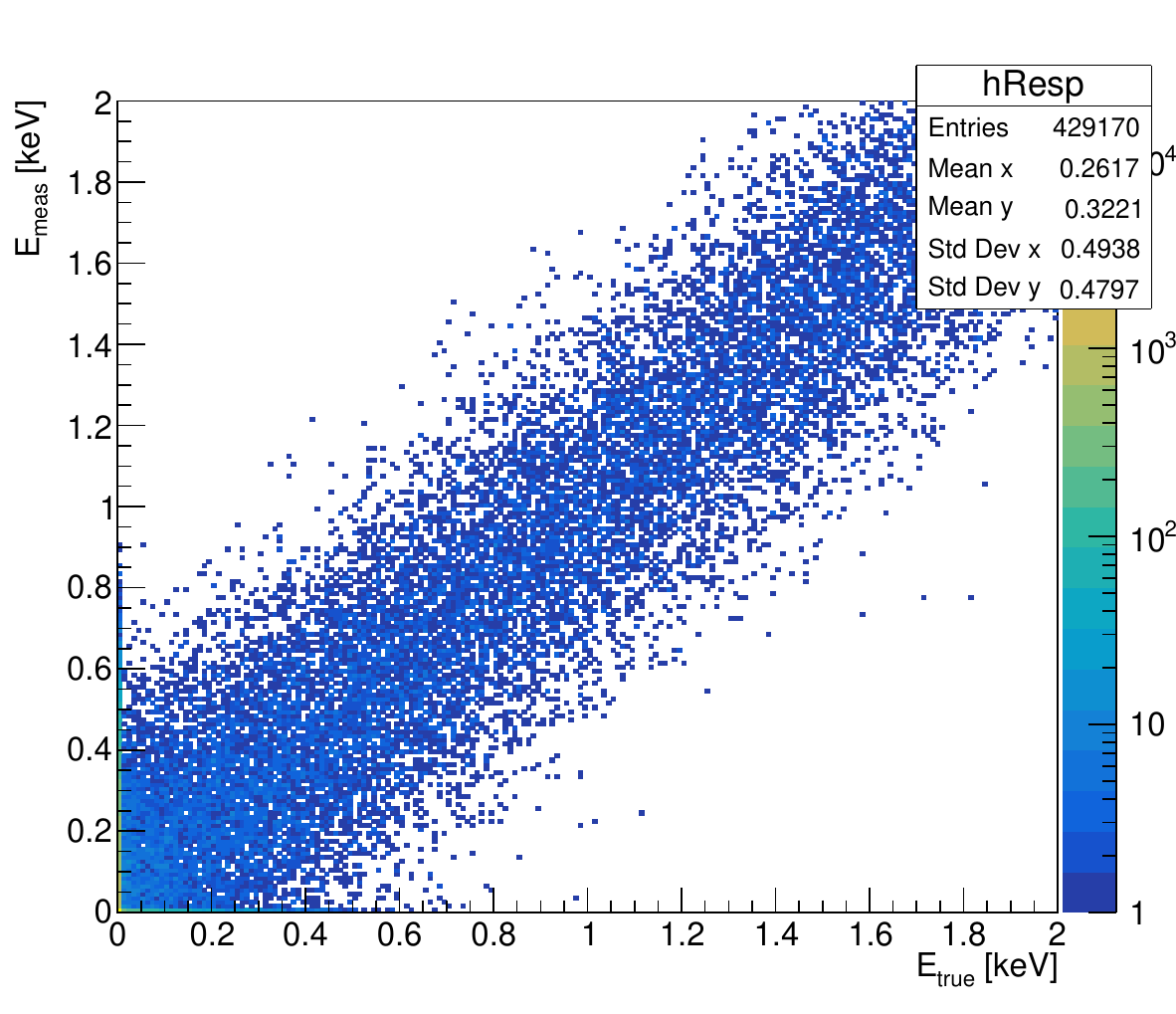}
    \caption{Titanium (Ti): detector response matrix.}
    \label{fig:matrix_Ti}
\end{subfigure}
\hfill
\begin{subfigure}{0.48\textwidth}
    \centering
    \includegraphics[width=\linewidth]{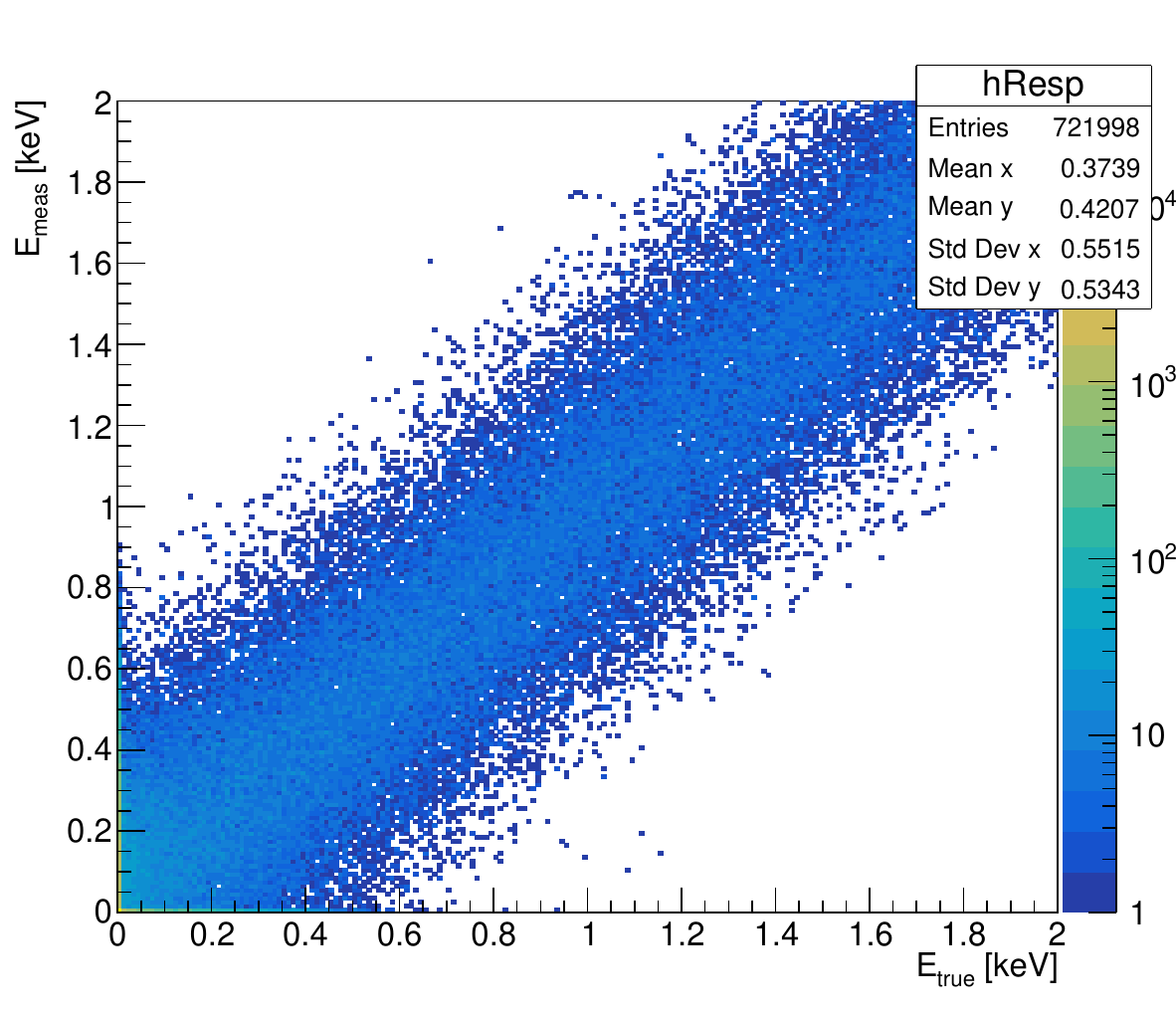}
    \caption{Zirconium (Zr): detector response matrix.}
    \label{fig:matrix_Zr}
\end{subfigure}

\caption{Detector response matrices mapping the generator-level (true) nuclear recoil energy, $E_{\mathrm{true}}$, to the reconstructed/measured recoil energy, $E_{\mathrm{meas}}$, after applying the common event selection  for each target nucleus. The color scale encodes the event density in the ($E_{\mathrm{true}}$, $E_{\mathrm{meas}}$) plane, thereby visualizing resolution smearing, threshold-induced truncations, and analysis-driven distortions that shape the experimentally accessible recoil spectrum.}
\label{fig:response_matrix_alltargets}
\end{figure}

\subsection{Detector Response Matrix: $E_{\mathrm{meas}}$ vs $E_{\mathrm{true}}$}

Figure~\ref{fig:response_matrix_alltargets} presents the detector response matrices for boron, magnesium, titanium, and zirconium targets, showing the correlation between the true nuclear recoil energy $E_{\mathrm{true}}$ and the reconstructed (measured) energy $E_{\mathrm{meas}}$. These matrices provide a direct visualization of energy smearing, reconstruction bias, and threshold effects under realistic detector conditions.

For all targets, the dominant structure follows the diagonal $E_{\mathrm{meas}} \approx E_{\mathrm{true}}$, confirming that the reconstruction algorithm preserves the overall energy scale without introducing large systematic bias. However, the width of the response band around the diagonal exhibits a clear dependence on nuclear mass.

For the lightest target (boron), the response matrix displays a visibly broader dispersion around the diagonal, particularly in the low-energy region ($E_{\mathrm{true}} \lesssim 0.6$~keV). This broadening reflects the combined impact of limited signal amplitude and detector resolution near threshold. A non-negligible fraction of low-energy recoils is reconstructed with significant downward or upward fluctuations, increasing bin-to-bin migration effects. Such behavior is expected when the recoil-energy scale overlaps strongly with the detector threshold region.

Magnesium shows a noticeably improved clustering around the diagonal compared to boron. The intermediate nuclear mass shifts the recoil spectrum to slightly higher energies, reducing the relative influence of threshold smearing. The response band becomes narrower and more linear over the 0.5--1.5~keV range, indicating improved reconstruction stability and reduced stochastic distortion.

For the heavier targets, titanium and especially zirconium, the response matrices exhibit the tightest correlation between $E_{\mathrm{true}}$ and $E_{\mathrm{meas}}$. The response band is significantly narrower, and deviations from linearity are minimal across the full visible energy range. This behavior indicates superior reconstruction fidelity and reduced sensitivity to detector-resolution effects. Although heavier nuclei produce lower maximum recoil energies for a fixed neutrino spectrum, the reconstructed energies are less dominated by threshold fluctuations, resulting in a more stable event migration pattern.

Another important feature is the asymmetry at very low $E_{\mathrm{true}}$. For all targets, the response density decreases sharply below the effective reconstruction threshold, reflecting the expected efficiency turn-on behavior. However, this cutoff is most pronounced for boron and least pronounced for zirconium, consistent with their respective efficiency curves.

From a detector-design perspective, these response matrices highlight a fundamental mass-dependent trade-off. Light nuclei offer extended kinematic recoil reach but suffer from stronger resolution-induced migration and threshold sensitivity. Heavy nuclei provide more robust energy reconstruction with reduced smearing, enhancing spectral stability and unfolding reliability.

The behavior observed for titanium and zirconium closely mirrors trends reported in current CEvNS experiments employing heavy or intermediate-mass targets (e.g., CsI and liquid argon detectors), where reconstruction robustness and controlled smearing are critical for precision cross-section measurements. The improved diagonal sharpness seen for the heavier targets in Figure~\ref{fig:response_matrix_alltargets} suggests that such materials may offer advantageous detector-level performance in future CEvNS implementations operating under realistic threshold conditions.

\begin{figure}[H]
\centering

{\Large \textbf{Selection efficiency curves as a function of $E_{\mathrm{true}}$}}\\[0.5cm]

\begin{subfigure}{0.48\textwidth}
    \centering
    \includegraphics[width=\linewidth]{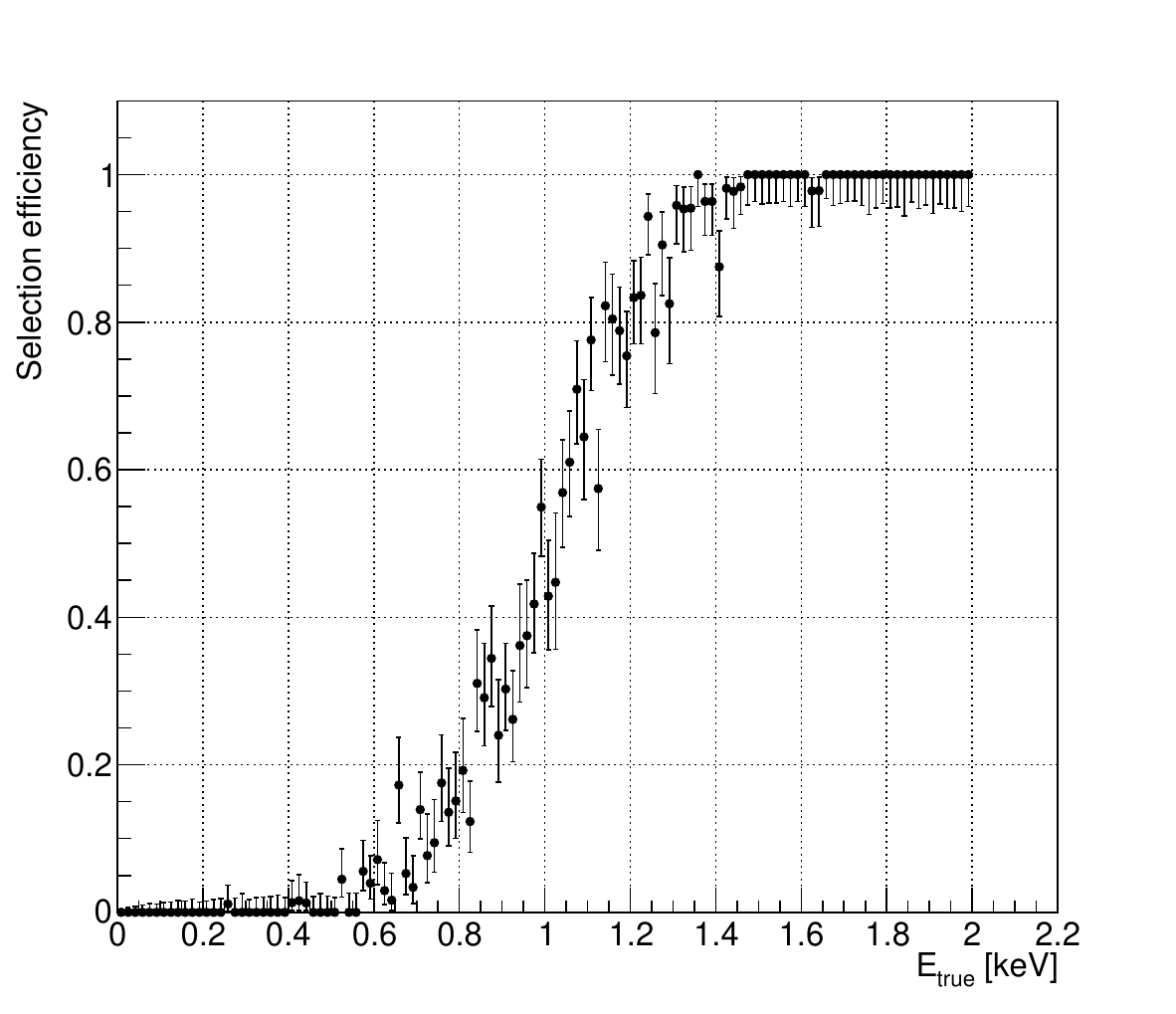}
    \caption{Boron (B) target.}
    \label{fig:eff_B}
\end{subfigure}
\hfill
\begin{subfigure}{0.48\textwidth}
    \centering
    \includegraphics[width=\linewidth]{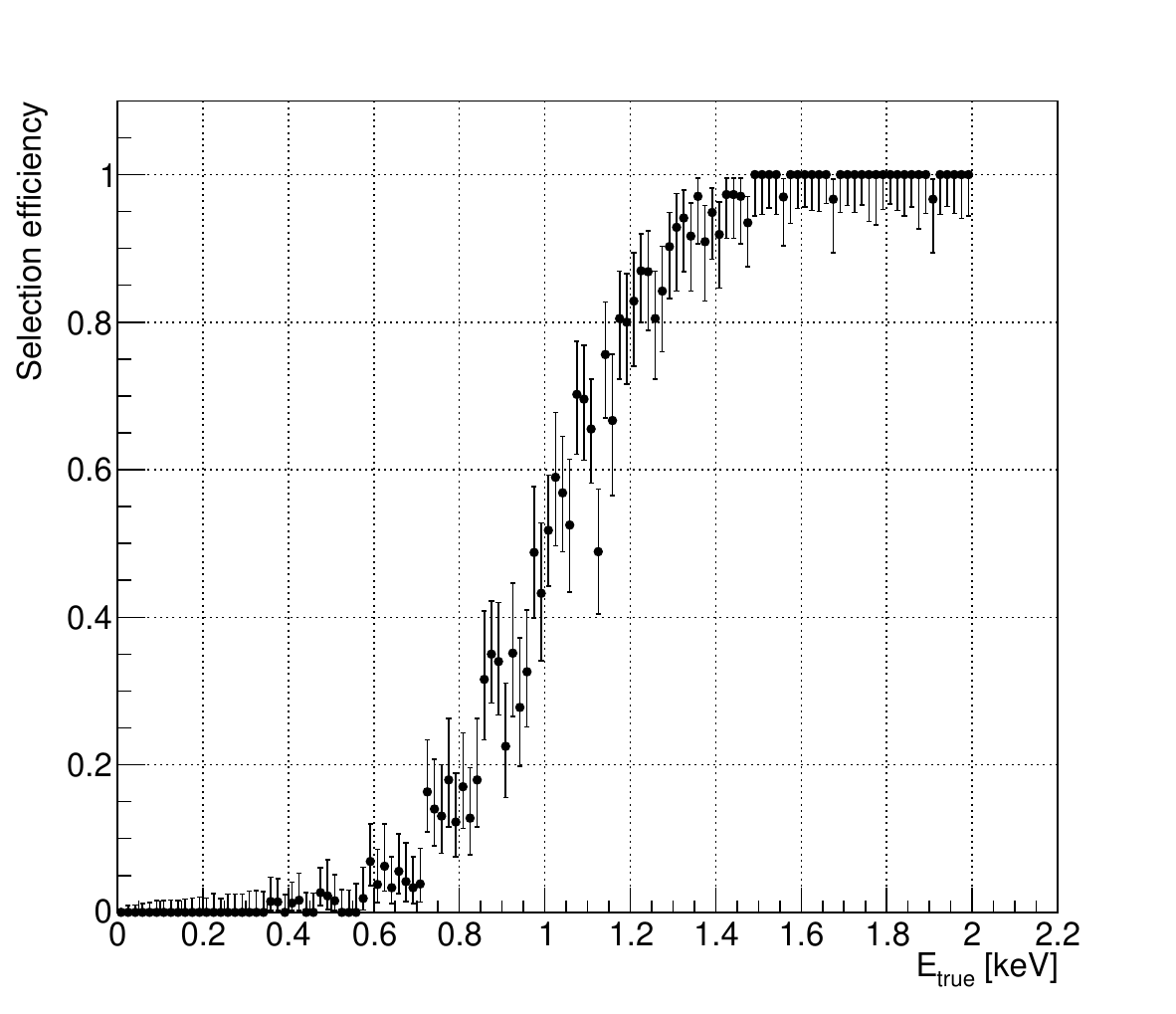}
    \caption{Magnesium (Mg) target.}
    \label{fig:eff_Mg}
\end{subfigure}

\vspace{0.5cm}

\begin{subfigure}{0.48\textwidth}
    \centering
    \includegraphics[width=\linewidth]{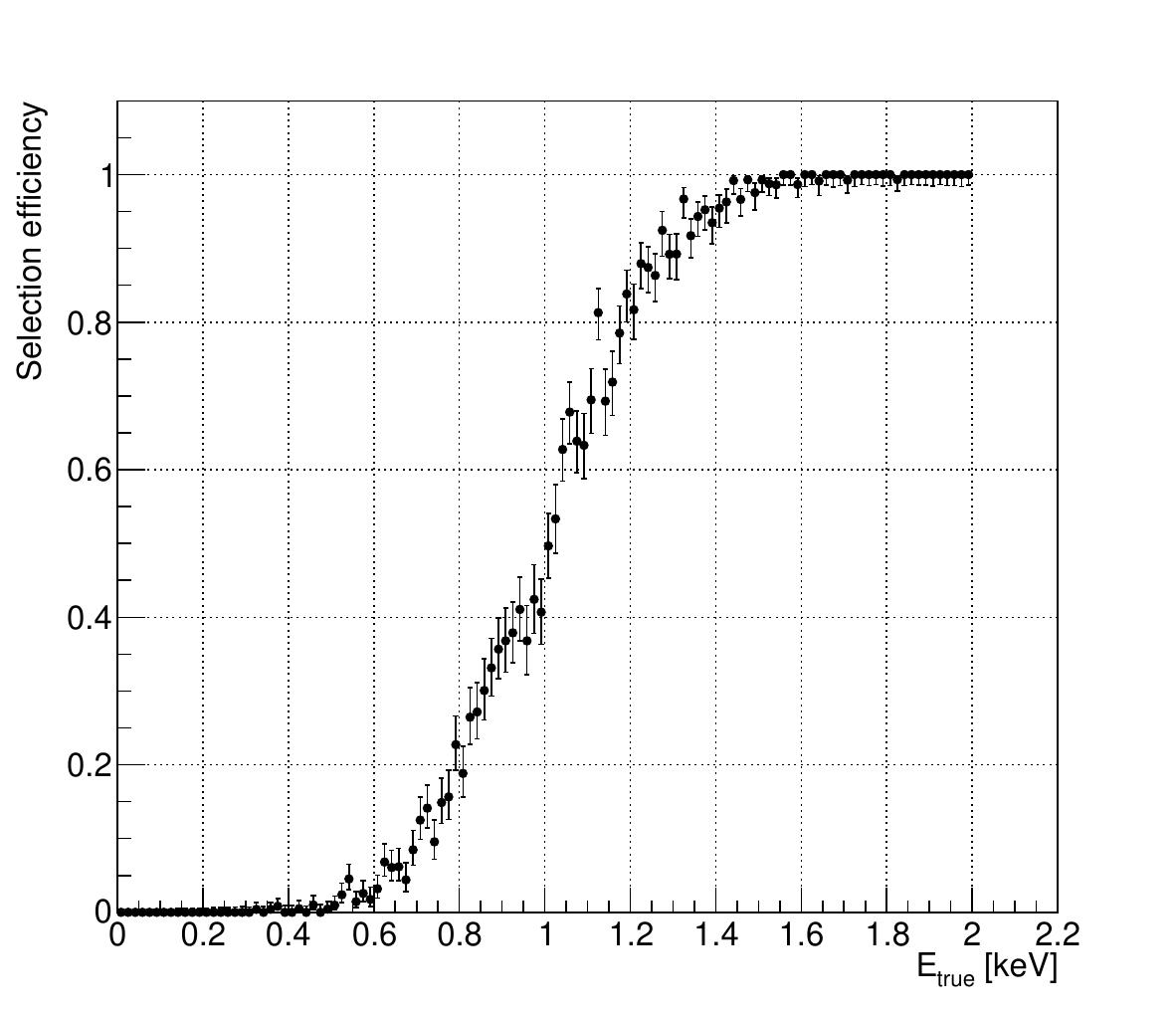}
    \caption{Titanium (Ti) target.}
    \label{fig:eff_Ti}
\end{subfigure}
\hfill
\begin{subfigure}{0.48\textwidth}
    \centering
    \includegraphics[width=\linewidth]{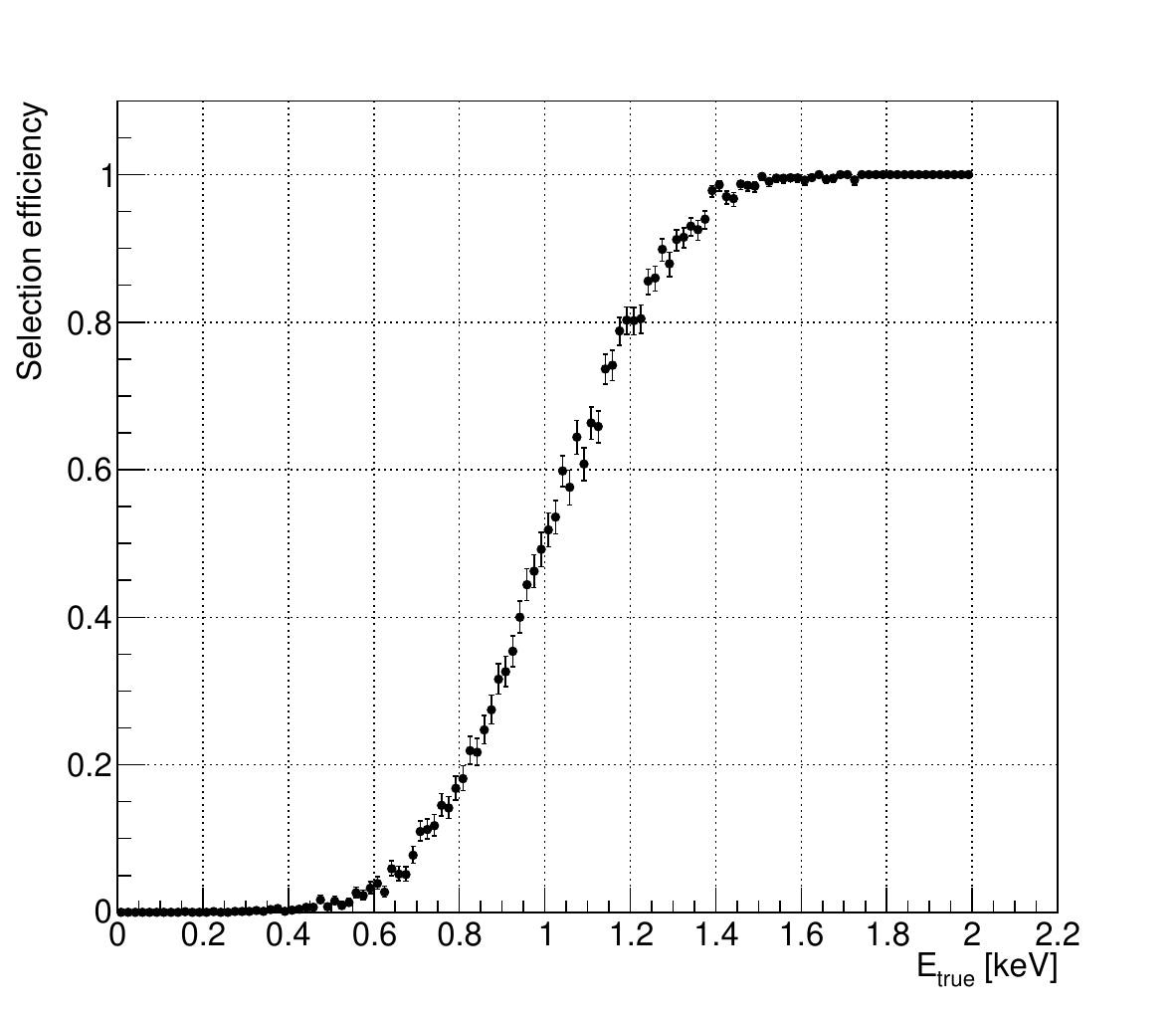}
    \caption{Zirconium (Zr) target.}
    \label{fig:eff_Zr}
\end{subfigure}

\caption{Selection efficiency $\epsilon(E_{\mathrm{true}})$ for the four target nuclei, defined as 
$\epsilon(E_{\mathrm{true}})=N_{\mathrm{reconstructed}}/N_{\mathrm{generated}}$ 
in bins of the true nuclear recoil energy. 
Here, $N_{\mathrm{generated}}$ denotes the number of Monte Carlo events produced in a given $E_{\mathrm{true}}$ bin, and $N_{\mathrm{reconstructed}}$ represents the subset of events successfully reconstructed by the detector-response model. 
No additional analysis-level cuts beyond the detector threshold and veto model described in Section 2.3 are applied.Error bars reflect finite Monte Carlo statistics in each $E_{\mathrm{true}}$ bin.}
\label{fig:efficiency}
\end{figure}

The selection efficiency curves shown in Figure~\ref{fig:efficiency} provide a compact and physically transparent summary of the detector-level acceptance as a function of the true nuclear recoil energy, $E_{\mathrm{true}}$. These curves encode the combined impact of trigger threshold, reconstruction resolution, and analysis-level selection criteria, and therefore directly determine the fraction of physically produced CE$\nu$NS events that remain observable after realistic detector effects are applied.

For all targets, the efficiency exhibits the characteristic ``turn-on'' behavior expected in low-threshold rare-event searches: $\epsilon(E_{\mathrm{true}}) \approx 0$ at very low recoil energies, followed by a rapid rise toward unity once the recoil signal exceeds the effective detection threshold. This behavior is a generic consequence of finite energy resolution and threshold cuts in CE$\nu$NS experiments, where the recoil spectrum is strongly concentrated in the sub-keV to few-keV region depending on the neutrino energy and nuclear mass.

A clear mass-dependent trend is observed across the four targets. Boron (Z=5), being the lightest nucleus considered, displays the slowest and most gradual efficiency turn-on. The lower recoil energies characteristic of light nuclei result in a larger fraction of events lying near or below the detection threshold, making the efficiency highly sensitive to resolution effects and bin-to-bin statistical fluctuations. Consequently, the efficiency curve for boron shows larger uncertainties in the threshold region and a broader transition band.

Magnesium (Z=12) exhibits a visibly steeper rise compared to boron, indicating improved matching between the recoil-energy distribution and the assumed detector threshold. The intermediate nuclear mass of magnesium shifts a larger fraction of the recoil spectrum into an experimentally accessible region, thereby reducing sensitivity to threshold smearing. This behavior reflects the well-known trade-off in CE$\nu$NS target optimization between recoil-energy reach and event-rate scaling.

For the heavier nuclei titanium (Z=22) and zirconium (Z=40), the efficiency turn-on becomes both steeper and more stable. Although heavier nuclei produce lower maximum recoil energies for a fixed neutrino energy, the increased coherent cross-section scaling approximately as $N^2$ enhances the overall event yield. As a result, the statistical robustness of the efficiency determination improves, and the transition to $\epsilon \simeq 1$ occurs more uniformly. In particular, zirconium demonstrates the most stable plateau behavior, with minimal fluctuations once above threshold, indicating strong resilience against detector-resolution effects.

This trend is consistent with current CE$\nu$NS experimental strategies. The COHERENT program has demonstrated that heavier targets such as CsI and liquid argon provide statistically robust signals despite relatively low recoil energies~\cite{COHERENT2017Science,COHERENT2022CsICrossSection}. Similarly, ongoing reactor-based efforts such as CONNIE emphasize the critical role of sub-keV nuclear-recoil acceptance and threshold control in determining overall physics reach. In all cases, the precise location and steepness of the efficiency turn-on region directly affect the extracted event rate and cross-section sensitivity.

From a detector-design perspective, these efficiency curves highlight a non-trivial optimization problem. Light targets offer access to higher recoil energies but suffer from stronger threshold-induced losses, whereas heavier targets provide enhanced coherence-driven rates but demand extremely low and stable energy thresholds to fully exploit the available signal. The results presented here quantitatively demonstrate how this balance manifests at the detector-response level, thereby providing a comparative framework for future target-material selection under realistic experimental conditions.

\section{Discussion}

The updated Monte Carlo statistics provide a significantly more stable and physically interpretable picture of detector-level CE$\nu$NS performance across the four target nuclei. 
When considered collectively, Figures~\ref{fig:etrue}, \ref{fig:emeas}, \ref{fig:response_matrix_alltargets}, and \ref{fig:efficiency} reveal a consistent mass-dependent structure that aligns with the theoretical expectations of coherent elastic neutrino--nucleus scattering.

\paragraph{Scope and modeling assumptions.}

The present study is designed as a detector-level and target-comparison analysis under controlled and identical simulation conditions. 
In order to isolate genuine target-mass and detector-response effects, a simplified neutrino source configuration is adopted. 
The emphasis of this work is therefore not on extracting absolute event rates or performing a flux-normalized sensitivity projection, 
but rather on investigating relative spectral distortions, reconstruction fidelity, and efficiency behavior across different nuclear targets.

A realistic reactor or spallation neutron source flux model can be folded with the response matrices presented here in a straightforward manner. 
Since all targets are simulated under the same neutrino energy configuration, the comparative conclusions regarding recoil-energy reach, 
detector smearing sensitivity, and efficiency turn-on behavior remain robust and flux-independent at the level of relative performance.

\paragraph{Background considerations.}

No explicit background model is included in the present analysis. 
This is a deliberate choice consistent with the scope of the study, which focuses on intrinsic detector-response properties 
and target-dependent recoil kinematics. 

The response matrices and efficiency curves derived here can be directly convolved with experiment-specific background spectra 
in future phenomenological applications. Therefore, the absence of an explicit background component does not compromise 
the validity of the target-comparison framework developed in this work.

It is important to emphasize that this work does not attempt to extract absolute sensitivity limits 
or cross-section measurements. Instead, it provides a controlled detector-level benchmarking framework 
that quantifies how target mass influences recoil observables under identical experimental assumptions. 

This approach enables future experimental studies to integrate realistic flux and background models 
while preserving the intrinsic response trends identified here.

\subsection*{True recoil spectra and nuclear-mass scaling}

The true recoil spectra (Figure~\ref{fig:etrue}) follow the characteristic CE$\nu$NS shape predicted by the Standard Model differential cross section
\cite{Freedman1974CEvNS,Scholberg2006Prospects}. 

For a fixed incident neutrino energy, heavier nuclei yield lower maximum recoil energies  for a given neutrino energy and neglecting flux convolution.

However, this reduction in endpoint energy is compensated by the coherent enhancement proportional to approximately $N^2$, 
which increases the overall interaction probability. 

In realistic neutrino spectra, the interplay between kinematics, flux distribution, and detector threshold 
determines the experimentally accessible recoil window. 
Therefore, the relative performance of light and heavy targets must always be interpreted 
in conjunction with the assumed neutrino-energy distribution.
 
This inverse scaling with nuclear mass is consistent with the recoil-energy relation
\[
E_r^{\max} \propto \frac{2E_\nu^2}{M},
\]
and reflects the trade-off between kinematic reach and coherent enhancement.

At the same time, the heavier targets benefit from the approximate $N^2$ scaling of the coherent cross section, leading to enhanced total event yield \cite{Freedman1974CEvNS}. 
This balance between spectral extent and total rate is a central design consideration in CE$\nu$NS experiments and motivates systematic target-material comparisons.

\subsection*{Measured spectra and detector-response effects}

The measured recoil spectra (Figure~\ref{fig:emeas}) show the expected distortion at low energies due to finite detector resolution and reconstruction effects. 
The deformation of the spectral shape near the lowest $E_{\mathrm{meas}}$ bins demonstrates how detector smearing modifies the physically produced spectrum, even in the absence of additional analysis-level selection cuts.

Heavier nuclei (Ti and especially Zr) exhibit more stable measured distributions, with a reduced relative impact of reconstruction fluctuations. 
This trend is qualitatively consistent with the performance of current CE$\nu$NS experiments employing medium-to-heavy nuclei such as CsI and liquid argon \cite{COHERENT2017Science,COHERENT2022CsICrossSection}. 
Although heavier targets produce lower maximum recoil energies, their statistically enhanced signal yields improve robustness against resolution-driven distortions.

\subsection*{Detector response matrices}

The detector response matrices (Figure~\ref{fig:response_matrix_alltargets}) provide the most direct diagnostic of reconstruction fidelity. 
For all targets, the dominant structure follows the expected diagonal correlation $E_{\mathrm{meas}}\simeq E_{\mathrm{true}}$, confirming the internal consistency of the response model.

However, the width of the response band varies with nuclear mass. 
Lighter nuclei exhibit broader dispersion at low energies, reflecting stronger relative sensitivity to reconstruction fluctuations. 
In contrast, titanium and zirconium display a tighter diagonal clustering and reduced reconstruction bias across the relevant energy range.

This behavior mirrors experimental experience: heavier targets used in CE$\nu$NS programs such as COHERENT demonstrate stable reconstruction performance despite reduced recoil endpoints \cite{COHERENT2021LArPRL,COHERENT2022CsICrossSection}. 
The response-matrix comparison therefore supports the interpretation that nuclear mass influences not only event statistics but also practical reconstruction stability.

\subsection*{Selection efficiency behavior}

The efficiency curves (Figure~\ref{fig:efficiency}), defined purely in terms of reconstruction success probability without additional veto or analysis cuts, exhibit the characteristic ``turn-on'' structure at low recoil energies. 
The steepness and stability of this rise differ among targets.

Heavier nuclei reach near-unity efficiency more rapidly and with smaller statistical fluctuations once the recoil energy exceeds the effective reconstruction scale. 
Lighter targets, while kinematically extending to higher $E_{\mathrm{true}}$, show larger bin-to-bin variations at low energies due to the reduced population in individual bins and enhanced sensitivity to reconstruction fluctuations.

This pattern is consistent with the detector-threshold challenges discussed in the CE$\nu$NS literature \cite{Scholberg2006Prospects,CONNIE2019DataSearch}. 
Although experiments such as CONNIE aim to exploit low-threshold silicon CCD technologies for low-energy recoils, maintaining stable efficiency at sub-keV scales remains experimentally demanding. 
Our detector-level comparison therefore quantitatively illustrates the intrinsic tension between low-energy reach and reconstruction robustness.

\section{Implications for Target Selection in Low-Threshold CEvNS Experiments}

The comparative results obtained in this study underscore that the selection of optimal target materials for Coherent Elastic Neutrino-Nucleus Scattering (CEvNS) cannot rely exclusively on theoretical cross-section scaling ($\propto N^2$) or intrinsic recoil kinematics. Instead, the viability of a target nucleus must be evaluated through a \textit{detector-aware framework} that explicitly accounts for the convolution of the physical recoil spectrum with realistic energy thresholds, efficiency turn-on curves, and reconstruction biases.

Our analysis demonstrates that while light nuclei such as boron and magnesium offer theoretically favorable kinematic endpoints, their experimental utility is severely constrained by the concentration of signal events in the sub-keV region, where detector efficiency is dominated by threshold effects and resolution smearing. In contrast, intermediate-mass nuclei, specifically titanium and zirconium, emerge as superior candidates that balance coherent enhancement with experimental robustness. These targets exhibit more stable detector response matrices and higher signal retention after selection cuts, thereby mitigating the systematic uncertainties associated with low-energy reconstruction. Consequently, these findings provide critical guidance for the design of next-generation CEvNS detectors, suggesting that maximizing physics reach requires prioritizing target materials that ensure spectral fidelity and reconstruction stability under realistic operating conditions.

\section{Conclusion}

In this work, we performed a detector-level comparative assessment of boron, magnesium, titanium, and zirconium as alternative CE$\nu$NS target materials under realistic experimental-response conditions.

Using high-statistics Monte Carlo simulations, we analyzed:
\begin{enumerate}
\item the true recoil-energy spectra,
\item the measured spectra after detector response,
\item the full response matrices $E_{\mathrm{meas}}$ vs.\ $E_{\mathrm{true}}$,
\item and the reconstruction efficiency curves.
\end{enumerate}

The results demonstrate a clear and systematic nuclear-mass dependence:

\begin{itemize}
\item Lighter targets extend recoil-energy reach but exhibit greater sensitivity to reconstruction fluctuations.
\item Heavier targets benefit from coherent enhancement and provide improved detector-level stability.
\end{itemize}

Among the studied materials, zirconium emerges as the most reconstruction-stable candidate within the considered configuration, while magnesium provides a balanced intermediate option between recoil reach and robustness.

Importantly, this study goes beyond purely theoretical cross-section comparisons by incorporating explicit detector-response modeling into the target-material evaluation. 
Such detector-level benchmarking is essential for realistic sensitivity projections and for guiding future CE$\nu$NS experimental strategies.

The framework developed here can be extended to additional target materials, refined response models, and experiment-specific configurations. 
As CE$\nu$NS measurements progress toward precision and new-physics sensitivity, systematic detector-level target optimization will become increasingly critical.

\section*{Acknowledgements}

The author acknowledges the use of the \textsc{Geant4} simulation toolkit and the \textsc{ROOT} data analysis framework in this study. 
Limited assistance from an AI-based language model (ChatGPT) was used exclusively for minor textual clarifications and formatting-related suggestions during manuscript preparation. 
All physics modeling, simulation setup, data analysis, and scientific interpretations presented in this work are entirely the responsibility of the author.

\nocite{*}
\bibliographystyle{unsrt}
\bibliography{cevns}

\end{document}